\documentclass[12pt]{article}
\usepackage[utf8]{inputenc}
\usepackage{amsmath}
\usepackage{amsfonts}
\usepackage{authblk}
\usepackage{multirow}
\usepackage{float}
\usepackage{graphicx}
\usepackage{booktabs}

\newcommand{\CS}{{\mathcal{C}}}
\newcommand{\be}{\begin{equation}}
\newcommand{\ee}{\end{equation}}
\newcommand{\baln}{\begin{eqnarray}}
\newcommand{\ealn}{\end{eqnarray}}
\newcommand{\ben}{\begin{equation*}}
\newcommand{\een}{\end{equation*}}

\newcommand{\eref}[1]{\eqref{#1}}

\title{Causal set d'Alembertians for various dimensions}
\author{Fay Dowker${}^1$}
\author{Lisa Glaser${}^2$}
\affil{${}^1$Blackett Lab, Imperial College, London SW7 2AZ, U.K.}
\affil{${}^2$The Niels Bohr Institute, Copenhagen University, \\ Blegdamsvej 17, DK-2100 Copenhagen \O, Denmark}
\date{\today}

\begin{document}

\maketitle

\begin{abstract}
We propose, for dimension $d$, a discrete Lorentz invariant operator on scalar fields that approximates the Minkowski spacetime scalar d’Alembertian.
  For each dimension, this gives rise to a
 scalar curvature estimator for causal sets, and 
 thence to a proposal for a causal set action. 
 
\end{abstract}

\section{\label{sec:dalemb}Discrete d'Alembertians}

In a discrete spacetime either locality or Lorentz invariance must be sacrificed. 
Causal sets are Lorentz invariant discrete spacetimes and are therefore
 fundamentally nonlocal. In order to base an approach to 
 quantum gravity on causal sets, one must confront this nonlocality and
 take up the challenge of recovering the local physics 
 that describes the physical world so well. A major step in this direction was 
 made when a discrete operator for scalar fields on causal sets
 was proposed that approximates the continuum d'Alembertian 
 when the causal set is well-approximated by 2 dimensional Minkowski spacetime
 \cite{Sorkin:2007qi, Henson:2006kf}.  
 This was extended to 4 dimensions \cite{Benincasa:2010ac} and we first review these
 2 and 4 dimensional operators. 
 
 Let $(\CS, \preceq)$ be a causal set 
 and let $\phi:\CS \rightarrow \mathbb{R}$
 be a scalar field on ${\mathcal{C}}$. For  $x,y \in \CS$ and such that 
 $y\preceq x$, 
 we define the inclusive  order interval $I(x,y): = \{ z\in \CS:\, y\preceq z\preceq x\}$ and
  $n(x,y): = | I(x,y) | -2$.\footnote{This is the convention for
$n(x,y)$ used in \cite{Sorkin:2007qi} and differs from that 
defined in \cite{Benincasa:2010ac}.}
  The 2 and 4 dimensional causal set d'Alembertians are, respectively \cite{Sorkin:2007qi,Benincasa:2010ac},
\begin{eqnarray}
B^{(2)}\phi(x)&:=&\frac{1}{l^2}\Big[- 2\phi(x)
+4\Big(\sum_{y \in L_1(x)}\!\!\!\phi(y)-
2\!\!\!\sum_{y\in L_2(x)}\!\!\!\phi(y)+\sum_{y\in L_3(x)}\!\!\!\phi(y)\Big)\Big]\,, \\
\label{Bop2}
B^{(4)}\phi(x)&:=&\frac{1}{l^2}\Big[-\frac{4}{\sqrt{6}}\phi(x)
+\frac{4}{\sqrt{6}}
\Big(\sum_{y \in L_1(x)} \!\!\!\phi(y)-
9\!\!\!\sum_{y\in L_2(x)}\!\!\!\phi(y)\nonumber\\ 
&&\ \ \ \ \ \ \ \ \ \ \ \ \ \ \ \ \ \ \ \ \ \ \ \ \ \ \ \ \ \ + 16\!\!\!\sum_{y\in L_3(x)}\!\!\!\phi(y)-8 \!\!\! \sum_{y\in L_4(x)}\!\!\!\phi(y)\Big)\Big]\,,
\label{Bop4}
\end{eqnarray}
where the sums run over the past ``layers'' $L_i(x):= \{ y\prec x:\, 
n(x,y)= i -1\}$. $l$ is a length, the analogue of the ``lattice spacing''.

Given a $d$ dimensional, causal Lorentzian manifold ${\mathcal{M}}$ with a scalar field $\phi:
{\mathcal{M}} \rightarrow \mathbb{R}$ of compact support and a point $x\in {\mathcal{M}}$, 
the operator $B^{(d)}$ gives rise to a  random variable, ${\mathbf{B}}^{(d)}\phi(x)$, in the following way. 
The Poisson process of {\textit{sprinkling}} \cite{Sorkin:2007qi} into 
${\mathcal{M}}$
at density $\rho = l^{-d}$ produces a random causal set $({\mathcal{C}}, \preceq)$
to which the point $x\in{\mathcal{M}}$ is added. $\phi$ restricts to a 
scalar field on the random causal set $\CS$ and 
the value of the random variable ${\mathbf{B}}^{(d)}\phi(x)$ is obtained by applying
 the operator $B^{(d)}$ to $\phi$ on $\CS$ at the point $x$. 
 Its expectation value 
in the sprinkling process  is denoted by
\begin{equation}
\bar{B}^{(d)}\phi(x) : =   {\mathbb{E}} \big({\mathbf{ B}}^{(d)}\phi(x)\big)\,.
\end{equation}
$\bar{B}^{(d)}\phi(x)$ depends on the manifold ${\mathcal{M}}$ and  $\rho = l^{-d}$ the sprinkling density, 
but we suppress these labels for ease of notation. 
When the Lorentzian manifold is $d$ dimensional Minkowski spacetime, in the limit of infinite 
density $\bar{B}^{(d)}\phi(x)$
tends to the continuum flat d'Alembertian of $\phi$ for $d=2$ \cite{Sorkin:2007qi}
and $d=4$ \cite{Benincasa:2010ac}:
\begin{equation}
\lim_{l \rightarrow 0} \bar{B}^{(d)}\phi(x)
= \Box^{(d)} \phi(x)\,.
\end{equation}

Following the form of the operator in 2 and 4 dimensions, consider the following ansatz for 
dimension $d$
\begin{equation}
B^{(d)}\phi(x)=\frac{1}{l^2}\left( \alpha_d \phi(x) +\beta_d \sum\limits_{i=1}^{n_d}  C^{(d)}_i \sum\limits_{y \in L_{i}}\phi(y) \right) \, ,
\end{equation}
where $n_d$ is the number of layers summed over,  and $\alpha_d$, $\beta_d$ and $C^{(d)}_i$ for
$ i = 1,\dots n_d$ are  constants to be determined. The first coefficient
$C^{(d)}_1$ is fixed to be
equal to $1$. 

As above, for a scalar field
$\phi$ of compact support on a $d$ dimensional 
Lorentzian manifold ${\mathcal{M}}$,  $x\in {\mathcal{M}}$ and a density $\rho = l^{-d}$
the Poisson process of sprinkling into ${\mathcal{M}}$ gives rise to a random variable
 ${\mathbf{B}}^{(d)}\phi(x)$, whose expected value 
 we denote by ${\bar{B}}^{(d)}\phi(x)$. 
 The Poisson distribution implies that this expected value is 
  \begin{eqnarray}
\bar{B}^{(d)} \phi(x)& =& \alpha_d{l^{-2}}  \phi(x)+ {\beta_d}{l^{-(d+2)}}  \!\! \int\limits_{J^-(x)} \!\! \mathrm{d}^dy \, \sqrt{-g(y)} \phi(y)\;   \nonumber \\ 
& & \ \ \ \ \ \ \ \ \ \ \ \ \ \ \ \ \ \ \ \ \  \times \sum\limits_{i=1}^{n_d} C^{(d)}_i \frac{ \left( V_d(y)l^{-d}\right)^{i-1}  }{(i-1)!}
 \exp(-V_d(y)l^{-d})\,.
\label{eqn:definition}
\end{eqnarray}
Here $J^-(x)$ is the causal past of $x$, and $V_d(y)$ is the spacetime volume of the causal interval between $x$ and $y$ in ${\mathcal{M}}$, {\textit{i.e.}} the intersection of the causal past of $x$ with the 
causal future of $y$. 
In the integral, $l^{-d} \sqrt{-g(y)} \mathrm{d}^dy $ is the probability that a point is 
sprinkled in the volume element at $y$ and the factor $\frac{ \left( V_d(y)l^{-d}\right)^{i-1}  }{(i-1)!}
 \exp(-V_d(y)l^{-d})$ is the probability that the element is in the $i$-th layer {\textit{i.e.} that there are 
exactly $i-1$ elements sprinkled in the interval between $x$ and $y$. 

If we define integrals
\begin{equation}
I_d(l):=\int\limits_{J^-(x)} \mathrm{d}^dy\,\sqrt{-g(y)}\,e^{-V_{d}(y)l^{-d}}\phi(y)
\end{equation}
it can be shown that 
\begin{eqnarray}
\bar{B}^{(2)} \phi(x)& =& \alpha_2{l^{-2}}  \phi(x)+ {\beta_2}{l^{-4}} {\cal{O}}_2 I_2(l)\\
\bar{B}^{(4)} \phi(x)& =& \alpha_4{l^{-2}}  \phi(x)+ {\beta_4}{l^{-6}} {\cal{O}}_4 I_4(l)
\label{eqn:ohat1}
\end{eqnarray}
where 
\begin{equation}
{\cal{O}}_2 = \frac{1}{8}(H + 2)(H+4)\ \ \ {\textrm{and}} \ \ \ {\cal{O}}_4 = \frac{1}{48}(H+2)(H+4)(H+6)
\label{eqn:ohat1}
\end{equation}
and
\begin{equation}
H = -l \frac{\partial}{\partial l}\,.
\end{equation}
The differential operator ${\cal{O}}_2$ annihilates $l^2$ and $l^4$
and thus eliminates
contributions to $\bar{B}\phi(x)$ that would not 
tend to zero in the $l \rightarrow 0$ limit. Our strategy will be to choose the 
appropriate differential operators ${\cal{O}}_d$ for general $d$
which will fix the constants $C^{(d)}_i$ via
\begin{equation}
{\cal{O}}_d \exp(-l^{-d} V) = \sum_{i =1}^{n_d} C^{(d)}_i \frac{(l^{-d} V)^{i-1}}{(i-1)!} \exp(-l^{-d} V)
\end{equation} 
and then to solve for $\alpha_d$ and $\beta_d$. 

We propose the following form of ${\cal{O}}_d$, for even dimensions $d=2n$, 
consistent with  \eref{eqn:ohat1},
\begin{eqnarray}\label{eq:op}
{\cal{O}}_{2n}=\frac{(H+2)(H+4) \dots (H+2n+2)}{2^{n+1}(n+1)!}\,,
\end{eqnarray}
and for odd dimensions, $d = 2n +1$, ${\cal{O}}_{2n+1} = {\cal{O}}_{2n}$. 

When the manifold is $d$ dimensional Minkowski spacetime, 
to evaluate the integrals we choose $x$ as the origin of 
coordinates and use radial null coordinates $v:=\frac{1}{\sqrt{2}}(t+r)$ and $u:=\frac{1}{\sqrt{2}}(t-r)$, where $r$ is the radius in $d-1$ dimensional spherical
coordinates centred on $x$. For $d>2$, the integral $I_d(l)$ is then
\begin{eqnarray}
I_d(l)=\int\limits^{0}_{-\infty} \mathrm{d}u \int\limits^{0}_{u} \mathrm{d}v \, r^{d-2} \int \mathrm{d}\Omega_{d-2} \, \phi(y) e^{-l^{-d} V_{d}(y)}  
\end{eqnarray}
where $\mathrm{d}\Omega_{d-2}$ is the integral over the $(d-2)$-sphere. 

\subsection{Three dimensions}
We will explicitly demonstrate the calculations in $3$ dimensional Minkowski spacetime, there the integral is:
\begin{eqnarray} \label{Igen3d}
I_3(l) =\int\limits^{0}_{-\infty} \mathrm{d}u \int\limits^{0}_{u} \mathrm{d}v \int\limits^{2 \pi}_0 \mathrm{d}\varphi \, \frac{1}{\sqrt{2}}(v-u) \phi(y) e^{-l^{-3} V_{3}(y)}  \, . 
\end{eqnarray}
The volume of the causal interval $V_{3}$ is
$\frac{\pi}{12}\tau^3 =\frac{\pi}{3\sqrt{2}} (u v)^{\frac{3}{2}}$. 
The operator in $3$d is \mbox{${\cal{O}}=\frac{1}{8}(H+2)(H+4)$}. This gives,
\begin{eqnarray} 
\bar{B}^{(3)} \phi(0)&:=&\alpha_3 l^{-2} \phi(0) +    \beta_3 \, l^{-5} \int\limits^{0}_{-\infty}\!\!\! \mathrm{d}u\!\! \int\limits^{0}_{u} \!\!\! \mathrm{d}v \!\!\int\limits^{2 \pi}_0 \!\mathrm{d}\varphi \frac{(v-u)}{\sqrt{2}} \phi(y)  \nonumber \\
 && \ \ \ \ \ \ \ \ \ \ \times \left( 1 - \frac{27}{8} {l'}^{-3} (u v)^{\frac{3}{2}} +\frac{9}{8} {l'}^{-6} (u v)^{3}  \right)  e^{-{l'}^{-3} (u v)^{\frac{3}{2}} } \label{eq:B3def}
\end{eqnarray}
where $l'$ is defined by $l^{-3} V_3(y) = {l'}^{-3} (uv)^{\frac{3}{2}}$. Note that $l'$ and $l$ are equal up to a factor 
of order one. 


The leading contribution to $\bar B$ as 
 $l \rightarrow 0$ comes from a neighbourhood of $x=0$, the 
origin of coordinates. This can be seen by analysing the integral (\ref{eq:B3def}). The integrand is 
exponentially suppressed for $uv \gg (l)^2$ and away from a neighbourhood of the origin, the 
integration range where the integral could possibly contribute in the limit is close to the light cone and
given by $ -L \le u \le -a $ and $ \frac{a^2}{u} \le v \le 0$ where $a>0$ is chosen to be much larger than 
$l$ and $L$ is a cutoff set by the size of the compact region of support of $\phi$. 
In the limit as $l$ gets small, $a$ can be chosen small enough that $\phi$ is approximately constant 
as a function of $v$ over this range of integration. The integral over $v$ at fixed $u$ of the 
integrand in (\ref{eq:B3def}) can be done and gives a dependence on $l$ which is a higher power than $l^5$ 
and so the contribution from close to the lightcone and away from the origin vanishes in the limit as $l$ gets small.

In this neighbourhood, defined by $ -a \le u,v \le 0$, the field can be Taylor expanded in Cartesian coordinates $\{y^\mu\}$
\begin{eqnarray}
\phi(y) &=& \phi(0)+y^\mu \partial_\mu \phi(0) +\frac{y^\mu y^\nu }{2}\partial_\mu \partial_\nu \phi(0)  + O(y^3) \, . %
\end{eqnarray}

Integrating over $\varphi$ gives:
\begin{eqnarray}
\int_0^{2\pi} \mathrm{d} \varphi   \phi(y) &=& 2\pi \phi(0) + \sqrt{2}\pi (u+v) \partial_t \phi(0) + \frac{\pi}{2}(u+v)^2 \partial_t^2\phi(0)  \nonumber \\
& & \ \ \ \ \ \ \ + \frac{\pi}{4}(v-u)^2 ( \partial_x^2 + \partial_y^2 ) \phi(0) + O_3(u,v)\,.
\end{eqnarray}
This can then be used in \eref{eq:B3def}, after which the integral can be done to give:
\begin{eqnarray}\label{eqn:givescontinuum}
\bar{B}^{(3)} \phi(0) &=& \alpha_3 l^{-2} \phi(0) +  \beta_{3} \bigg( l^{-2} \phi(0) +  \nonumber \\
&& \ \ \ \ \ \left(\frac{3\sqrt{2}}{\pi}\right)^{\frac{2}{3}} \Gamma\left(\frac{5}{3}\right) (- \partial^2_t + \partial^2_x +\partial^2_y)\phi(0) \bigg) + \dots
\end{eqnarray}
The corrections represented by $\dots$ can be obtained by calculating integral \eref{eq:B3def} with the lower limit of integration in $u$ set to $-a$. The correction terms that are first and second order in derivatives of $\phi$ are
\begin{eqnarray}
E(l,a)&=&
 -\left( \frac{3 \sqrt{2}}{\pi}\right)^{\frac{1}{3}} \frac{\sqrt{2}\; \partial_t \phi(x)}{ l^{-1} a^2 \;\Gamma\!\left(\frac{5}{3}\right)}  +\left(\frac{3 \sqrt{2}}{\pi}\right)^{\frac{1}{3}} \frac{\partial^2_t\phi(x)}{ l^{-1}  a \;\Gamma\!\left(\frac{5}{3}\right)} \nonumber \\
		& & \ \ \ \ \ \ \ \ \ \ \ -\left(\frac{3 \sqrt{2}}{\pi}\right)^{\frac{1}{3}} \frac{3 \;(\partial^2_x +\partial^2_y) \phi(x)}{ 2\; l^{-1} a\; \Gamma\!\left(\frac{5}{3}\right)} \,.\label{eq:corr}
\end{eqnarray}
There are also smaller corrections proportional to higher derivatives of $\phi$ as 
well as terms that go like $\exp(- \frac{\pi}{3\sqrt{2}} a^3 l^{-3})$, which are exponentially small for large $a l^{-1}$. From (\ref{eqn:givescontinuum}) we can read off,
$-\alpha_3 = \beta_3 =\left(\frac{\pi }{3\, \sqrt{2} }\right)^{2/3}\Gamma\left(\frac{5}{3}\right)^{-1}$, and then 
\begin{equation}
	\lim_{l \rightarrow 0} \bar{B}^{(3)}\phi(0)=\Box^{(3)} \phi(0)\,.
\end{equation}

\subsection{Other dimensions}
The calculation can be repeated for Minkowski spacetime  in other dimensions.
${\cal{O}}_d$ is given by (\ref{eq:op}) 
for $d$ even, and  ${\cal{O}}_{2m+1} = {\cal{O}}_{2m}$ so the number of layers is 
 $n_{d} = \frac{d}{2}+2$ for even 
dimensions and $n_{d} = \frac{d-1}{2} + 2$ for odd dimensions. 
The resulting $C^{(d)}_i$ for various dimensions are given in table \ref{tab:factors}. The case of one dimension is included for completeness and is the usual discretization of the second derivative. 

\begin{table}
\begin{center}
\begin{tabular}
{  l  c c c c c } \toprule
&$C_1$	&	$C_2$	&	$C_3$	&	$C_4$ &	$C_5$  \\ \midrule
1d &	$1$&$-\frac{1}{2}$&&&\\
2d &$1$&$-2$&$1$&&\\
3d&$1$&$-\frac{27}{8}$&$\frac{9}{4}$&&	\\
4d&$1$&$-9$&$16 $&$-8$&	\\
5d&$1$&$- \frac{215}{16}$&$\frac{225}{8}$&$-\frac{125}{8}$&\\
6d&$1$&$-34$&$141$&$-189$&$81$\\
7d&$1$&$-\frac{6307}{128}$&$\frac{14749}{64}$&$-\frac{10633}{32}$&$\frac{2401}{16}$\\ \bottomrule
\end{tabular}
\caption{Coefficients in the operator $B^{(d)}$}
\label{tab:factors}
\end{center}
\end{table}

The coefficients $\alpha_d$ and $\beta_d$ are given in table \ref{tab:vol+alphabeta}. 
The volume, $V_{0\,d(u,v)}$ of the flat causal interval in $d$ dimensions between $x$ at 
the origin of coordinates and the point $y$ with radial null coordinates $(u,v)$
is also given in table \ref{tab:vol+alphabeta}.

In general
\begin{equation}
V_{0\,d}(u,v) = c_d (uv)^{d/2} = S_{d-2} \frac{2^{\frac{2-d}{2}}}{d(d-1)} (uv)^{d/2}
\end{equation}
where $S_{d-2}$ is the volume of the $(d-2)$-sphere. 

For odd  $d$,  $\alpha_d= - c_d^{2/d}/\Gamma\!\left(\frac{d+2}{d}\right)$
while for even $d$ an additional factor of $2$ leads to $\alpha_d=-2c_d^{2/d}/\Gamma\!\left(\frac{d+2}{d}\right)$.  In 4 dimensions this simplifies to $\alpha_4 = -\frac{4}{\sqrt{6}}$. 

  $\alpha$ and $\beta$ are related by 
\begin{equation}
\frac{\alpha_d }{\beta_d}= - \lim_{l \rightarrow 0}
l^{-d}  S_{d-2}  {\cal{O}}_d \int_{-a}^0 \mathrm{d}u \int_u^0  \mathrm{d}v 
\left[\frac{v-u}{\sqrt{2}}\right]^{d-2} 
e^{-l^{-d} V_{0\,d(u,v)}}\,.
\end{equation}

\begin{table}
\begin{center}
\begin{tabular}
{  l  c c c } \toprule
Dim& $V_{{0}\, d} (u,v)$	&	$\alpha_{d}$	&	$\beta_{d}$ \\ \midrule
1d &	$\sqrt{2}(u v)^{1/2}$                                                        & $-1 $           & $2$\\
2d &	$(u v)^{2/2} $                                             & $-2$            &$4$\\
3d&$\frac{\pi}{3\sqrt{2}}(u v)^{3/2}$    &   $-\frac{1}{\Gamma\left(\frac{5}{3}\right)}
 \left(\frac{\pi }{3\, \sqrt{2}}\right)^{2/3}$&   $	\frac{1}{ \Gamma\left(\frac{5}{3}\right)}\left(\frac{\pi }{3\, \sqrt{2} }\right)^{2/3}	$    \\
4d&$\frac{\pi}{6}(u v)^{4/2}$                        &    $-\frac{2}{\Gamma\left(\frac{6}{4}\right)}\left(\frac{\pi}{6}\right)^{2/4}$             &    $ \frac{2}{\Gamma\left(\frac{6}{4}\right)}\left(\frac{\pi}{6}\right)^{2/4}$     \\
5d&$\frac{\pi^2}{20\sqrt{2}} (u v)^{5/2}$ &   $-\frac{1}{\Gamma\left(\frac{7}{5}\right)} \left(\frac{\pi^2}{20\sqrt{2}}\right)^{2/5}$              &   $\frac{3}{8 \; \Gamma\left(\frac{7}{5}\right)} \left(\frac{\pi^2}{20\sqrt{2}}\right)^{2/5}$	     \\
6d&$\frac{\pi^2}{45}(u v)^{6/2}$                   &    $-\frac{2}{\Gamma\left(\frac{8}{6}\right)}\left(\frac{\pi^2}{45}\right)^{2/6}$             &    $ \frac{4}{5\;\Gamma\left(\frac{8}{6}\right)}\left(\frac{\pi^2}{45}\right)^{2/6}$     \\
7d&$\frac{\pi^3}{168\sqrt{2}}(u v)^{7/2}$ &      $-\frac{1}{ \; \Gamma\left(\frac{9}{7}\right)}\left(\frac{\pi^3}{168\sqrt{2}} \right)^{2/7}$           &       $\frac{1}{8 \; \Gamma\left(\frac{9}{7}\right)}\left(\frac{\pi^3}{168\sqrt{2}} \right)^{2/7}$  \\ \bottomrule
\end{tabular}
\caption{Volumes of $d$-dimensional flat causal intervals and 
coefficients $\alpha_d$ and $\beta_d$}
\label{tab:vol+alphabeta}
\end{center}
\end{table}
\section{Nonlocality Scale}

For $d=2$ and $d=4$ the value of $\bar{B}^{(d)}\phi$ approaches 
its limiting value -- the continuum flat d'Alembertian -- when the 
discreteness scale is small enough that the field is slowly varying on this 
scale, however for each individual sprinkling the fluctuations around 
the mean value are enormous and the fluctuations {\textit{grow}} with increasing 
sprinkling density. We expect the same behaviour in other dimensions. 
This makes numerical simulations difficult and
in order to tame the fluctuations, Sorkin introduced a family of 
operators  for $d = 2$  parametrised by a fixed 
physical ``nonlocality scale'' which is larger than the 
discreteness scale.  Each operator in the family gives 
approximately the same mean 
as $B^{(2)}$, so long as the field is slowly varying 
on the nonlocality scale but the fluctuations around the mean  
now diminish as the discreteness scale tends to zero \cite{Sorkin:2007qi}. 
This introduction of a nonlocality scale
was  extended to $d=4$ \cite{Benincasa:2010ac}\footnote{There is an error in  \cite{Benincasa:2010ac} where, using the 
convention defined in that paper, $n(x,y)$  should be replaced  by 
$n(x,y) -2$ in the formula for the nonlocal operator.} and we here give the family of 
nonlocal operators in $d$ dimensions. Recall 
 \begin{equation}
B^{(d)}\phi(x)=\frac{1}{l^2}\left( \alpha_d \phi(x) +\beta_d \sum\limits_{i=1}^{n_d}  C^{(d)}_i \sum\limits_{y \in L_{i}}\phi(y) \right) \, .
\end{equation}
Let $\xi$ be the nonlocality length scale such that $\xi\ge l$,  and define $\epsilon :=\left( \frac{l}{\xi}\right)^d$. 
Define, for each $d$, a one parameter family of operators on scalar fields on a causal set 
 \begin{equation}
B^{(d)}_\epsilon \phi(x):=\frac{\epsilon^{\frac{2}{d}}} {l^2}\left( \alpha_d \phi(x) +
\beta_d \epsilon \sum\limits_{y\prec x}f_d(n(x,y),\epsilon)\phi(y)\right) \, , \label{eq:nonlocdAlembertian}
\end{equation}
where the sum is over elements $y$ in the causal set to the past of $x$ and
\begin{equation}\label{eq:lazybones}
f_d(n,\epsilon ):=(1-\epsilon)^n
\sum_{i = 1}^{n_d} C_i^{(d)} {n \choose i-1} \left(\frac{ \epsilon}{1-\epsilon}\right)^{i-1}\,.
\end{equation}

As before, the sprinkling process at 
density $\rho = l^{-d}$ into a $d$ dimensional 
spacetime with a  marked point $x$ and a
scalar field turns this operator into a random variable whose
mean, $\bar{B}^{(d)}_\epsilon \phi(x)$ has the same form as equation (\ref{eqn:definition}) 
 but with $l$ replaced by $\xi$: 
\begin{eqnarray}\label{eq:bbar_epsilon}
\bar{B}^{(d)}_\epsilon \phi(x)& =& \alpha_d{\xi^{-2}}  \phi(x)+ {\beta_d}{\xi^{-(d+2)}}  \!\! \int\limits_{J^-(x)} \!\! \mathrm{d}^dy \sqrt{-g(y)} \phi(y)\;   \nonumber \\ 
& & \ \ \ \ \ \ \ \ \ \ \ \ \ \times \sum\limits_{i=1}^{n_d} C^{(d)}_i \frac{ \left( V_d(y)\xi^{-d}\right)^{i-1}  }{(i-1)!}
 \exp(-V_d(y)\xi^{-d})\,.
\end{eqnarray}
To see that this is the case, first note that the first term in (\ref{eq:nonlocdAlembertian}) is constant in the 
sprinkling process
and so its mean is equal to it and this gives the first term in (\ref{eq:bbar_epsilon}). 
To derive the integral we look at each term in (\ref{eq:lazybones}). 
Consider the  random variable
\begin{equation*}
C_i^{(d)} \epsilon \sum_{y \prec x} (1 - \epsilon)^{n(x,y)}  {n(x,y) \choose i-1} \left(\frac{ \epsilon}{1-\epsilon}\right)^{i-1} \phi(y)\,.
\end{equation*}
Its mean over sprinklings  is 
\begin{eqnarray*}
C_i^{(d)} \epsilon  \,\rho \int d^dy& \sqrt{-g(y)} \,\phi(y)\\
 &\sum_{n=i-1}^\infty (1 - \epsilon)^n  {n \choose i-1} \left(\frac{ \epsilon}{1-\epsilon}\right)^{i-1} \frac{(\rho V_d(y))^n}{n!} e^{-\rho V_d(y)}\,,
\end{eqnarray*}
where $\rho \,d^dy \sqrt{-g(y)}$ is the probability that an element is sprinkled in an element of volume at $y$ and 
$\frac{(\rho V)^n}{n!} e^{-\rho V}$ is the probability that exactly $n$ elements are sprinkled into 
a region of volume $V$. Using $\epsilon\rho = \xi^{-d}$, this becomes
\begin{eqnarray*}
C_i^{(d)} \epsilon  \,\rho \int d^dy \sqrt{-g(y)} \,\phi(y)  \epsilon^{i -1} \frac{(\rho V_d(y))^{i -1}}{(i-1)!} e^{-\epsilon \rho V_d(y)} \\
=C_i^{(d)} \xi^{-d} \int d^dy \sqrt{-g(y)} \,\phi(y)  \frac{ (\xi^{-d}V_d(y))^{i -1}}{(i-1)!} e^{-\xi^{-d} V_d(y)} \, ,
\end{eqnarray*}
hence the result.

\subsection{Simulations}

Numerical simulations of $\bar{B}^{(3)}_\epsilon \phi(x)$ were done for small causal sets
sprinkled into 3 dimensional Minkowski spacetime. The sprinkling region is a causal interval 
defined by $-L \le u\le 0$ with the cutoff parameter $L=1$.
The test functions used were $\phi(x)=1$, $\phi(x)= x^2$  and $\phi(x)= t^2$ and they were measured on causal sets with a mean number of elements $\langle N \rangle =1000$. 
 The discreteness length $l$ is set by $l^3 =\frac{1}{1000} \frac{\pi}{3 \sqrt{2}}$ 
and the nonlocality length $\xi$ is parametrised by $\epsilon^{\frac{1}{d}} = l/\xi$ which was varied between values
$0.1$ and $0.9$. 
The graph in Figure (\ref{simulationresults}) plots the mean over $100\; 000$ sprinklings of $B_{\textrm{corr}}\phi$,
where $B_{\textrm{corr}}\phi$ is defined as the discrete operator $B_{\epsilon}\phi$ minus the correction term
$E(\xi,L)$. The correction $E(\xi,L)$ is defined in equation (\ref{eq:corr}) and arises as a
boundary effect at finite sprinkling density:
\begin{equation}
	B_{\textrm{corr}}\phi:=B_{\epsilon}\phi -E(\xi,L)
\end{equation}
For the simulations it is important that we subtract the corrections, defined in equation (\ref{eq:corr}), before comparing to the continuum prediction, since our causal sets are quite small and the corrections are non-negligible. 

\begin{figure}
\includegraphics[width=\linewidth]{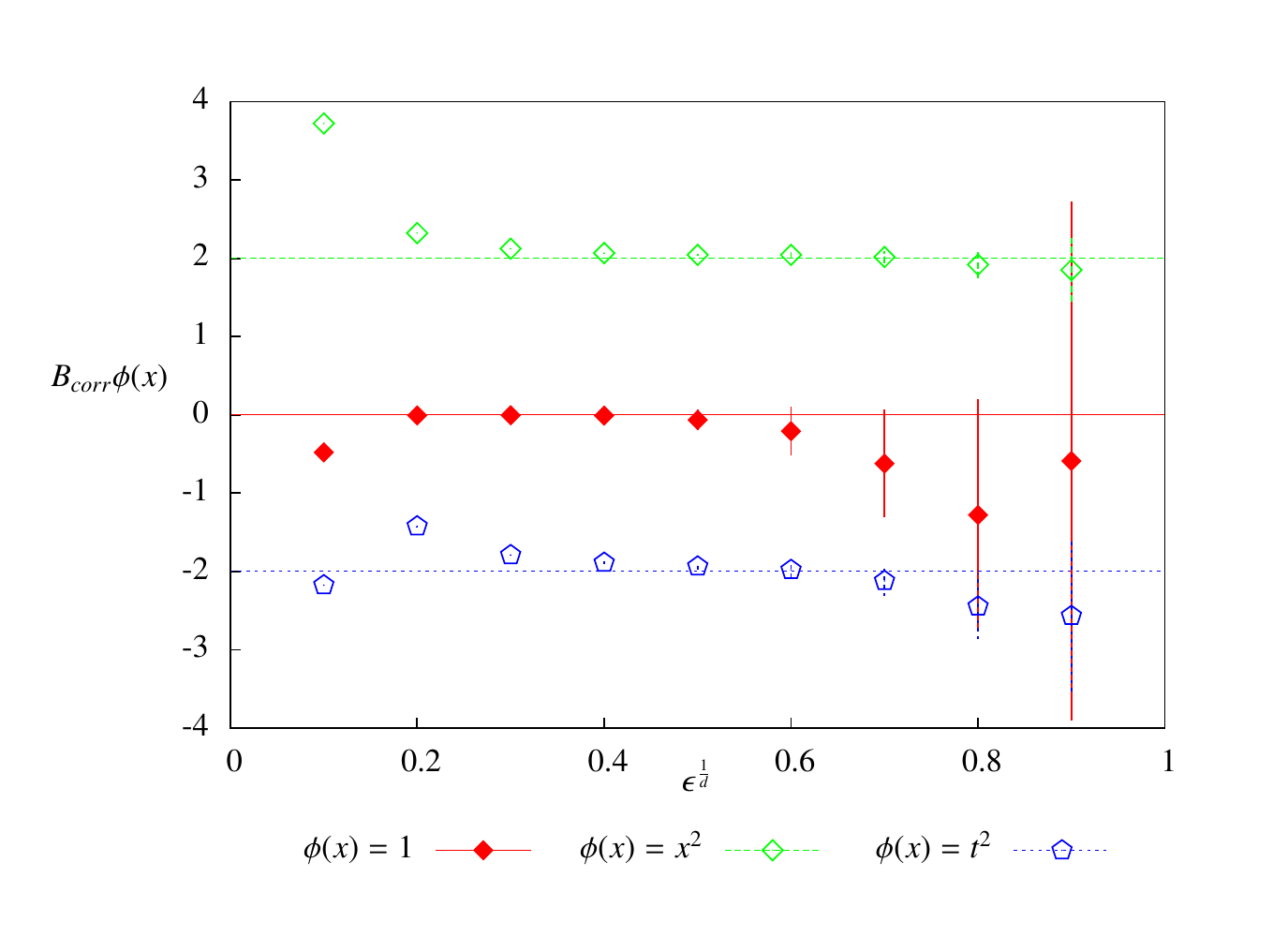}
\caption{\label{simulationresults} $B_{\textrm{corr}}\phi$ plotted against $\epsilon^{\frac{1}{d}}$. The error bars 
show the standard error.}
\end{figure}

The data are consistent with the expected result. The best results are achieved with $\epsilon^{\frac{1}{d}}$
between $0.4$ and $0.5$. For epsilon close to one, the nonlocality length is approximately the 
discreteness length and the fluctuations in the results are large, as expected. For small $\epsilon$ the nonlocality length becomes of order the cutoff length and the deviations from the continuum 
d'Alembertian have two origins. One is that the sprinkled region is not large enough to accommodate the 3 ``layers'' that give the necessary cancellations and the second is that we have ignored the
 ``exponentially small'' corrections which are becoming non-negligible for these parameter values. 
For the causal sets used in the simulations the linear 
size is only about ten times the discreteness length and the good approximation to 
the continuum values is evidence that causal sets can efficiently encode 
geometrical information. 

\section{Scalar Curvature}

Having identified the coefficients in the operator $B^{(d)}$ we now investigate the 
mean, $\bar{B}^{(d)}\phi$, of the random variable when the manifold, ${\mathcal{M}}$, 
is a $d$ dimensional curved Lorentzian manifold:
\begin{equation}\label{eqn:curved}
\bar{B}^{(d)} \phi(x) = \alpha_d{l^{-2}}  \phi(x)+ {\beta_d}{l^{-(d+2)}}  
{\mathcal{O}}_d \!\! \int\limits_{J^-(x)} \!\! \mathrm{d}^dy \, \sqrt{-g(y)} \phi(y) \exp(-V_dl^{-d})\,, \; 
\end{equation}
where $V_d := V_d(y)$, the volume of the causal interval between $x$ and $y$. 

We assume that the spacetime region of compact support of the field is small 
compared to the radius of curvature of the spacetime. Then all curvature corrections to 
the flat space integral will be small and we can assume that the mean tends to a local limit as $l\rightarrow 0$ 
as it does in the flat space case. Then 
by dimensional arguments the limit will be a linear combination of $\Box^{(d)} \phi(x)$ and $R(x)\phi(x)$ where 
$\Box^{(d)} = g^{\mu\nu}\nabla_\mu\nabla_\nu$ is the d'Alembertian 
and $R$ is the Ricci scalar curvature of $\mathcal{M}$:
\begin{eqnarray}\label{eq:boxricci}
\lim_{l \rightarrow 0} 
{\bar{B}}^{(d)}\phi(x) &=& \Box^{(d)} \phi(x) + a_d R(x)\phi(x)\,,
\end{eqnarray}
where $a_d$ is a dimensionless constant. 
For $d=2$ and $d=4$, the value of the coefficient $a$ is $-\frac{1}{2}$ \cite{Benincasa:2010ac}. 
We seek to determine $a_d$ for dimension $d$. 

To obtain the local result (\ref{eq:boxricci}), the integral (\ref{eqn:curved}) is performed over
 a normal neighbourhood of $x$,
$\mathcal{N}$, with Riemann normal coordinates, $\{y^\mu\}$, centred at $x$ as the origin of coordinates. 
 $\sqrt{-g}$ and $V_d$ can be expanded to first order in the curvature:
\begin{eqnarray}
 \sqrt{-g} &=& 1-\frac{1}{6} R_{\mu \nu}y^\mu y^\nu + O(R^2) = 1+\delta \sqrt{-g} + O(R^2)\,, \\ 
 V_{d} &=&V_{0\, d} \bigg( 1- \frac{d\, R \eta_{\mu\nu} y^\mu y^\nu}{24 (d+1)(d+2)} 
+ \frac{d \,R_{\mu \nu} y^\mu y^\nu}{24 (d+1)} + O(R^2) \bigg)  \nonumber \\
& =& V_{{0}\, d}+ \delta V_d + O(R^2) \; ,\label{eq:deltas}
\end{eqnarray}
where $V_{0\,d}$ is the volume 
of the causal interval in $d$ dimensional flat space between the origin and 
a point with Cartesian coordinates  $\{y^\mu\}$ and all curvature components are evaluated at the origin. 
The formula (\ref{eq:deltas}) is from \cite{Volumes} where the curvature components are evaluated at 
the centre of the causal interval between the origin and $y$: to this order in $R$, the formulae are the
same.

Expanding \eref{eqn:curved} in powers of curvature we get
 \begin{equation}\label{eqn:expand}
\bar{B}^{(d)} \phi(x) = \bar{B}_0^{(d)} \phi(x) + \delta\bar{B}^{(d)}\phi(x)  + O(R^2).
\end{equation}
The zeroth order term is exactly the Minkowski space integral and will give
$\Box^{(d)} \phi(x)$ in the limit because the scalar d'Alembertian at $x$ in Riemann normal 
coordinates is $\eta^{\mu\nu}\partial_\mu\partial_\nu$.

To determine the coefficient $a_{d}$ it is sufficient to consider 
$\phi(x) = 1$ and in that case the term first order in curvature is \begin{eqnarray}
\delta\bar{B}^{(d)}&\,:=&  \beta l^{-(d+2)} {\mathcal{O}}_d \int\limits_{{\mathcal{N}}} \!\! \mathrm{d}^dy
 \bigg( \delta\sqrt{-g}  - l^{-d} \delta{V}	\bigg) e^{- l^{-d} V_{0\,d}}	\;  \\
&\,=&\beta l^{-(d+2)}  {\mathcal{O}}_d \int\limits_{{\mathcal{N}}} \!\! \mathrm{d}^dy \bigg( -\frac{1}{6} R_{\mu \nu}y^\mu y^\nu			 \nonumber	\\ 				
& &\ \ \ \ \ \ \ \ \ \ \ - l^{-d}  V_{0} \bigg\{- \frac{d}{24 (d+1)(d+2)}R \,\eta_{\mu\nu} y^\mu y^\nu \nonumber\\
& &\ \ \ \ \ \ \ \ \ \ \ \ \ \ \ \ + \frac{d}{24 (d+1)}R_{\mu\nu} y^\mu y^\nu \bigg\}\bigg)	 e^{-l^{-d} V_{0}}\; . 
\end{eqnarray}

Letting $t := y^0$ and $r^2 := \sum_i (y^i)^2$ and transforming to spherical polar coordinates, 
we can integrate over the angular coordinates. $V_{0\,d}$ depends only on $t$ and $r$ and
is rotationally symmetric so 
the only terms in $R_{\mu \nu} y^\mu y^\nu$ that remain are the diagonals $R_{00} t^{2}$ and $ R_{ii} r^{2} $. 
Since we are operating in Riemann normal coordinates  we can use $\Sigma_{i} R_{ii}=(R_{00}+R)$ 
to get rid of $R_{ii}$.
The result is
\begin{eqnarray}
&& \beta l^{-(d+2)} S_{d-2} {\mathcal{O}}_d\int_{\mathcal{N}} \!\! \mathrm{d} t\, \mathrm{d}r \, r^{d-2} 
  \nonumber \\
&& \times \bigg[ \bigg( R_{00} t^2 + \frac{1}{d-1}(R_{00}+ R) r^2\bigg)
             \bigg(-\frac{1}{6} - \frac{1}{24(d+1)}l\frac{\partial}{\partial l}\bigg) e^{-l^{-d}V_{0\,d}}
	\nonumber	\\ 				
             && \ \ \ \ \ \ \ \ \ \ \ \ \ \ \ \ + R(t^2 - r^2)\frac{1}{24(d+1)(d+2)} l  \frac{\partial}{\partial l} e^{-l^{-d}V_{0\,d}} \bigg]\,.
\end{eqnarray}
We can then split this into an integral multiplying $R_{00}$ and a different integral multiplying $R$, which can be solved independent of each other.
In lightcone coordinates, the integral multiplying $R_{00}$ is 
\begin{eqnarray}
I^{(d)}_{00}&: = & {\mathcal{O}}_d \bigg(-\frac{1}{6} - \frac{1}{24(d+1)}l\frac{\partial}{\partial l}\bigg) \int_{-a}^0 {\mathrm{d}}u  \!\! \int_{u}^0 \mathrm{d} v \, \bigg(\frac{v-u}{\sqrt{2}}\bigg)^{d-2} \nonumber \\
& & \ \ \ \ \ \ \ \ \ \ \ \ \ \  \ \ \  \times \bigg( \frac{(v+u)^2}{2}+ \frac{(v-u)^2}{2(d-1)}\bigg)
            e^{-l^{-d}V_{0\,d}}\,,
\end{eqnarray}
and the coefficient multiplying $R$ is
\begin{eqnarray}
I^{(d)}_R &:= & \beta l^{-(d+2)} S_{d-2} {\mathcal{O}}_d  \int_{-a}^0 {\mathrm{d}}u  \!\! \int_{u}^0 \mathrm{d} v \, 
\bigg(\frac{v-u}{\sqrt{2}}\bigg)^{d-2}
 \nonumber \\
& & \times \bigg(  \frac{(v-u)^2}{2(d-1)}\bigg(-\frac{1}{6} - \frac{1}{24(d+1)}l\frac{\partial}{\partial l}\bigg) e^{-l^{-d}V_{0\,d}}  \nonumber \\
           & &\ \ \ \ \ \ \ \ \ \ \ \ \ \ \ \ \ \ + 2 uv  \frac{1}{24(d+1)(d+2)} \; l \frac{\partial}{\partial l}  e^{-l^{-d}V_{0\,d}} \bigg)\,.
\end{eqnarray}

The volume $V_{0\,d}$ is given as a function of $u$ and $v$ in table \ref{tab:vol+alphabeta} 
for dimensions $d=2$ to $7$ and direct calculation of the integrals  in these cases
gives that $\lim_{l \rightarrow 0} I^{(d)}_{00}=0$
and $\lim_{l \rightarrow 0} I^{(d)}_R = - \frac{1}{2}$ for each of these dimensions. So, 
in these dimensions, if there is a local 
limit, then 
\begin{equation}\label{coeffhalf}
\lim_{l \rightarrow 0} \bar{B}^{(d)} \phi(x) =\Box^{(d)} \phi(x)  -\frac{1}{2} R(x)  \phi(x) \;.
\end{equation}

We make the obvious conjecture that the result (\ref{coeffhalf}) holds
 in all dimensions $d$.

As in 2 and 4 dimensions \cite{Benincasa:2010ac}, these results about 
the d'Alembertian operator in other dimensions give rise to proposals for 
scalar curvature estimators and actions for causal sets. 
For a causal set $\CS$ and $x \in \CS$ we define
\begin{equation}
\mathcal{R}^{(d)}(x) = -\frac{2}{l^2} \bigg( \alpha_d + \beta_d \sum_{i = 1}^{n_d} C_i ^{(d)} N_i(x) \bigg)\,,
\end{equation}
where $N_i(x)$ is the cardinality of the $i$th layer $L_i(x)$ to the past of $x$. 
When evaluated on causal sets sprinkled into a $d$ dimensional Lorentzian manifold, 
this becomes a random variable whose mean will under appropriate 
conditions be close to the continuum scalar curvature $R(x)$.
Summing $\mathcal{R}^{(d)}(x)$ over all elements of the causal set, gives the following 
proposal for an action for the causal set, 
\begin{equation}
\frac{1}{\hbar}\mathcal{S}^{(d)}(\CS) = \zeta_d \bigg[ N + \frac{\beta_d}{\alpha_d}
 \sum_{i = 1}^{n_d} C_i ^{(d)} N_i \bigg]\,,
 \end{equation}
where $\zeta_d = - \alpha_d \left(\frac{l}{l_p}\right)^{{d-2}}$ and $l_p^{d-2} = 8 \pi G\hbar$. 
The case $d=2$  has been studied for flat regions of Minkowski spacetime, a
cylinder spacetime and the topology  changing 
trousers \cite{Benincasa:2010as}.

The nonlocal discrete d'Alembertian, \eref{eq:nonlocdAlembertian} gives rise, similarly, to the 
nonlocal action,
\begin{equation}
\frac{1}{\hbar}\mathcal{S}_\epsilon^{(d)}(\CS) = \zeta_d \bigg[ \epsilon^{\frac{2}{d}}N + 
\epsilon^{\frac{d+2}{d}}\frac{\beta_d}{\alpha_d} \sum_{x\in \CS}\sum_{y\in \CS \atop y \prec x} 
f_d\left( n(x,y), \epsilon\right) \bigg]\,.
\end{equation}
 For the random variable  we expect fluctuations around the mean to be dampened by the 
nonlocal averaging.

\section{Summary}
This paper generalises the work of \cite{Sorkin:2007qi, Henson:2006kf,Benincasa:2010ac} 
from 2 and 4 dimensions to $d$ dimensions. We gave a
a general procedure to obtain a scalar d'Alembertian operator for causal sets in any dimension. 
We have shown numerical evidence that the operator approximates the flat space scalar
d'Alembertian in 3 dimensions. In curved spacetime,  when the mean 
of the operator on a scalar field $\phi$ has a local limit as the discreteness 
length tends to zero, the limit 
will be $\Box^{(d)} - \frac{1}{2} R$, where $R$ is the scalar curvature, 
for $d = 2,3,\dots 7$. It would be 
good to prove the conjecture for all $d$. We used the  discrete d'Alembertian to 
propose causal set actions in all dimensions. The operators and actions have
nonlocal versions which give a way 
to damp the fluctuations about the mean. It would be
interesting to explore these actions using Monte Carlo
simulations of the path integral \cite{Surya:2011du}. 

\section{Acknowledgements}
We thank D. Benincasa, S. Johnston and  B. Schmitzer, for helpful discussions. Special thanks go to N. Hustler who did the simulations. FD thanks the Perimeter Institute for Theoretical Physics for support. Research at Perimeter Institute is supported by the Government of Canada through Industry Canada and by the Province of Ontario through the Ministry of Economic Development and Innovation. 
\bibliographystyle{unsrt}
\bibliography{refs}

\begin{thebibliography}{1}

\bibitem{Sorkin:2007qi}
Rafael~D. Sorkin.
\newblock {Does locality fail at intermediate length-scales?}
\newblock In D.~Oriti, editor, {\em {Approaches to Quantum Gravity: Towards a
  New Understanding of Space and Time}}. Cambridge University Press, 2006.

\bibitem{Henson:2006kf}
Joe Henson.
\newblock {The causal set approach to quantum gravity}.
\newblock In D.~Oriti, editor, {\em {Approaches to Quantum Gravity: Towards a
  New Understanding of Space and Time}}. Cambridge University Press, 2006.

\bibitem{Benincasa:2010ac}
Dionigi M.~T. Benincasa and Fay Dowker.
\newblock {The Scalar Curvature of a Causal Set}.
\newblock {\em Phys. Rev. Lett.}, 104:181301, 2010.

\bibitem{Volumes}
G.~W. Gibbons and S.~N. Solodukhin.
\newblock {The geometry of small causal diamonds}.
\newblock {\em Phys. Lett.}, B649:317--324, 2007.

\bibitem{Benincasa:2010as}
Dionigi~M.T. Benincasa, Fay Dowker, and Bernhard Schmitzer.
\newblock {The Random Discrete Action for 2-Dimensional Spacetime}.
\newblock {\em Class.Quant.Grav.}, 28:105018, 2011.

\bibitem{Surya:2011du}
Sumati Surya.
\newblock {Evidence for a Phase Transition in 2D Causal Set Quantum Gravity}.
\newblock {\em Class.Quant.Grav.}, 29:132001, 2012.

\end{thebibliography}

\end{document}